\RequirePackage{fix-cm}
\documentclass[smallextended]{svjour3}       % onecolumn (second format)
\smartqed  % flush right qed marks, e.g. at end of proof
\usepackage{graphicx}
\usepackage{hyperref}

\usepackage{color,soul}

\renewcommand\hl[1]{#1}

\journalname{Trans. Indian Nat. Acad. Eng.}
\begin{document}

\title{Atomistic simulations of twin boundary effect on the crack growth behaviour in BCC Fe}

\author{G. Sainath \and   A. Nagesha }


\institute{G. Sainath \at
              Scientific Officer-E, Materials Development and Technology Division, Metallurgy and Materials Group, Indira Gandhi Centre for Atomic Research, Kalpakkam, Tamilnadu-603102, India, Tel.: 044-27480500-21213\\
              \email{sg@igcar.gov.in}  %  \\
           \and
           A. Nagesha \at
              Scientific Officer-G \& Head, Fatigue Studies Section, Materials Development and Technology Division, Metallurgy and Materials Group, Indira Gandhi Centre for Atomic Research, HBNI, Kalpakkam, Tamilnadu-603102, India \\
              }

\date{Received: / Accepted: }

\maketitle

\begin{abstract}
In this paper, the effect of twin boundaries on the crack growth behaviour of single crystal BCC Fe has been investigated using 
molecular dynamics simulations. The growth of an atomically sharp crack with an orientation of (111)$<$110$>$ (crack plane/crack 
front) has been studied under mode-I loading at constant strain rate. In order to study the influence of twin boundaries on the 
crack growth behaviour, single and multiple twin boundaries were introduced perpendicular to crack growth direction. The results 
indicate that the (111)$<$110$>$ crack in single crystal BCC Fe grows in brittle manner. However, following the introduction of 
twin boundaries, a noticeable plastic deformation has been observed at the crack tip. Further, increasing the number of twin 
boundaries increased the amount of plastic deformation leading to better crack resistance and high failure strains. Finally, an 
interesting relationship has been observed between the crack growth rate and flow stress.\\

%\PACS{PACS code1 \and PACS code2 \and more}
%\subclass{MSC code1 \and MSC code2 \and more}
\end{abstract}

\section{Introduction}
\label{intro}
Twin boundaries (TBs) were special class of boundaries with lowest interface energy and high degree of symmetry. TBs 
contain arrangements of atoms that are mirror reflections of those on the other side, separated by twin plane. The low 
energy of TBs results in a number of superior properties relative to conventional grain boundaries (GBs). For example, 
it has been found that the TBs enhance the strength without loss of ductility \cite{lu2009revealing,li2010dislocation}, 
improve fracture toughness and crack resistance \cite{qin2009enhanced,singh2011fracture,liu2014atomistic}, improve 
corrosion resistance and strain rate sensitivity \cite{deng2010effects}. Moreover, the TBs posses high thermal and 
mechanical stability \cite{anderoglu2008thermal,wang2013structure}. Due to these unique properties, the materials 
containing high density of TBs attracted a lot of interest in materials community.

In the past, many studies have been carried out to understand the strengthening and softening mechanisms due to the 
presence of TBs in FCC \cite{lu2009revealing,li2010dislocation} and BCC \cite{sainath2016deformation,xu2017size,xu2018deformation} 
materials. Further, the role of TBs on the crack propagation behaviour in FCC materials is also well understood 
\cite{liu2014atomistic,zhu2011modeling}. It has been shown that the TBs in FCC materials enhance the crack growth 
resistance due to different dislocation-twin and crack-twin interactions. For example, molecular dynamics (MD) 
simulations in twinned Cu have shown that the dislocations nucleated from the crack tip accumulate at the TBs and 
produce strain hardening which results in high plastic deformation and crack blunting \cite{zhu2011modeling}. TBs 
also deflect the crack propagation path in Ag due to change in slip plane orientation across the boundary \cite{liu2014atomistic}. 
The periodic deflection of crack path by TBs result in fracture with zigzag cracks in nanotwinned Cu \cite{zeng2015fracture}. 
Apart from crack blunting and deflection, the crack arrest and crack closure have also been observed due to twinning or 
de-twinning process occurring at the TBs \cite{liu2014atomistic,zhou2015atomistic,kim2012situ}. Further, during the 
crack growth in twinned Ag, an alternating blunting mechanism has been reported \cite{liu2014atomistic}. The crack 
is blunted when it is away from the TBs and becomes sharp closer to the TBs \cite{liu2014atomistic}. Cheng et al. 
\cite{cheng2010intrinsic} have investigated the inter-granular fracture along the TBs in Cu using MD simulations. 
A directional anisotropy is observed in crack propagation in terms of brittle cleavage in one direction and dislocation 
emission in the opposite direction \cite{cheng2010intrinsic}. In contrast to numerous studies in FCC materials, the role 
of TBs on the crack propagation behaviour in BCC materials has not been investigated particularly at the atomic scale. 
In view of this, the present study is aimed at understanding the role of TBs on the crack propagation behaviour in BCC Fe.

\section{MD Simulation details}
\label{Details}

All MD simulations have been carried out in Large scale Atomic/Molecular Massively Parallel Simulator (LAMMPS) package 
\cite{plimpton1995fast} employing an embedded atom method (EAM) potential for BCC Fe given by Mendelev and co-workers 
\cite{mendelev2003development}. The Mendelev potential has been widely used to study the deformation and fracture behaviour 
of BCC Fe \cite{sainath2016deformation,terentyev2013blunting,kedharnath2017molecular,sainath2017atomistic,sainath2016directional}. 
The visualization of atomic configurations has been performed by AtomEye \cite{li2003atomeye} and OVITO 
\cite{stukowski2009visualization} using centro-symmetry parameter (CSP) \cite{kelchner1998dislocation}.

\begin{figure}[h]
\centering
\includegraphics[width=9cm]{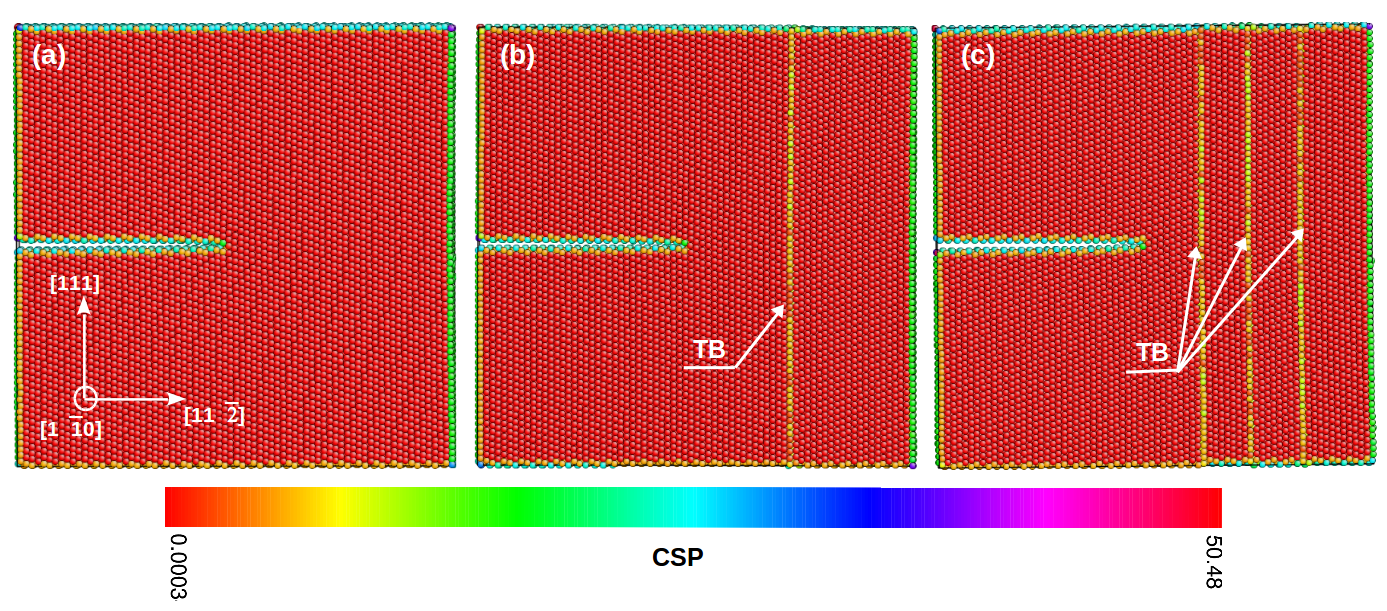}
\caption {\small The initial simulation models considered in this study. The atomically sharp (111)$<$110$>$ crack in (a) 
perfect BCC Fe, (b) sample containing a single TB and, (c) sample containing multiple TBs. The atoms are coloured according 
to centro-symmetry parameter (CSP) \cite{kelchner1998dislocation} \hl{and the color legend of CSP is shown in the bottom for 
reference}.}
\label{Initial}
\end{figure}

Initially, a single crystal BCC Fe oriented in $[1\bar10]$, $[11\bar2]$ and [111] crystallographic directions has been 
created. Following this, an atomistically sharp crack was introduced on (111) plane with $[1\bar10]$ being the crack 
front direction i.e. the (111)$<$110$>$ crack (Figure \ref{Initial}a). The specimen had the dimensions of 1.6 nm $\times$ 
17.3 nm $\times$ 17.3 nm. The crack length was half the width of the specimen i.e., 8.65 nm. In order to investigate the 
influence of TBs on the crack growth behaviour, single and multiple TBs were introduced perpendicular to the plane of 
crack as shown in Figure \ref{Initial}b-c. The periodic boundary conditions were applied only in the crack front direction, 
$[1\bar10]$. The model systems was equilibrated to a temperature of 10 K in NVT ensemble \hl{with an integration time step 
of 2 fs}. Upon completion of equilibrium process, the growth of an atomically sharp (111)$<$110$>$ crack has been studied in 
perfect (Figure \ref{Initial}a) as well twinned BCC Fe (Figure \ref{Initial}b-c) under mode-I loading \hl{employing the same NVT 
ensemble and time step}. The loading has been applied at a constant engineering strain 
rate of $1\times10^8$ s$^{-1}$ by imposing velocity to atoms along the [111] axis that varied linearly from zero at the 
bottom \hl{fixed layer} to a maximum value at the top layer. The average stress is calculated from the Virial expression 
\cite{zimmerman2004calculation}.

\section{Results and Discussion}
\label{Results}

\begin{figure}
\centering
\includegraphics[width=5cm]{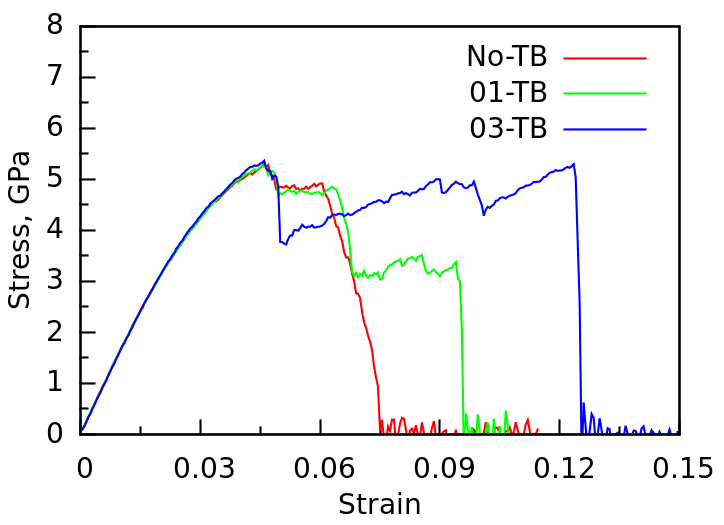}
\caption {\small The stress-strain behaviour of pre-cracked perfect BCC Fe along with that of BCC Fe containing one and three 
twin boundaries perpendicular to the crack growth direction.}
\label{Stress-strain}
\end{figure}

Figure \ref{Stress-strain} shows the stress-strain behaviour of pre-cracked perfect BCC Fe along with that of specimen 
containing one and three TBs perpendicular to the crack growth direction. It can be seen that during the elastic deformation, 
the stress-strain behaviour in all the samples is almost similar up to the peak value of stress. Following the elastic 
deformation, the flow stress shows a gradual drop from the peak value, which indicates the onset of crack growth. The 
stress drop is marginal in the case of perfect as well as sample with one TB, while significant drop is observed in sample 
with three TBs (Figure \ref{Stress-strain}). Following the drop, the flow stress varies from sample to sample mainly due to 
their different internal microstructure, i.e., TBs. In perfect sample, the flow stress is constant up to certain value of 
strain and then rapidly falls to zero indicating the failure at a strain value of 0.075 (Figure \ref{Stress-strain}). On the 
other hand, the flow stress in a sample containing a single TB shows an interesting step-wise decreasing behaviour, where 
the constancy over a period of strain is followed by a drop. Finally, it fails at a strain value of 0.10, which is relatively 
higher than the failure strain of perfect sample (Figure \ref{Stress-strain}). In contrast, the flow stress of the sample 
containing three TBs shows a two-step gradually increasing behaviour before dropping to zero at a strain value of 0.122 
(Figure \ref{Stress-strain}). These results clearly show that increasing the number of TBs increases the plastic deformation 
and failure strain of BCC Fe. 

\begin{figure}
\centering
\includegraphics[width=9cm]{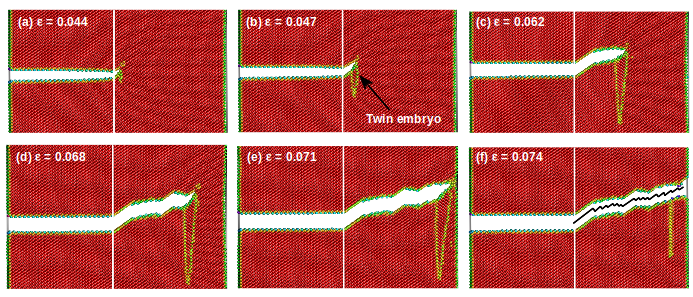}
\caption {\small The atomic snapshots displaying the growth of (111)$<$110$>$ crack in perfect BCC Fe as a function of strain.
\hl{The coloring of atoms and the color legend is same as in Fig. 1. The viewing direction is $<$110$>$ and only the portion of (110) 
plane is shown.}}
\label{Perfect}
\end{figure}

In order to understand the mechanisms behind the role of TBs in increasing the failure strain of BCC Fe, the atomic configurations 
have been analyzed for pre-cracked perfect as well as twinned BCC Fe. Figure \ref{Perfect} shows the sequence of atomic snapshots 
displaying the growth of (111)$<$110$>$ crack in perfect BCC Fe without any TBs. It can be seen that the crack growth is initiated 
by the de-cohesion of atomic bonds on \{110\} plane (Figure \ref{Perfect}a) followed by the nucleation of a micro-twin at the crack 
tip, which is orientated at $90^o$ from the crack plane (Figure \ref{Perfect}b). With increasing strain, the crack changes its path 
from a single \{110\} plane to a combination of a small steps on two different \{110\} planes, which is effectively a \{112\} plane 
(Figure \ref{Perfect}c). During this crack growth, many partial dislocations nucleate from the crack tip and glide on TBs leading to 
the growth as well as migration of a micro-twin (Figure \ref{Perfect}c). This growth and migration of micro-twin is responsible for 
constant flow stress observed in the strain range 0.045 - 0.06 (Figure \ref{Stress-strain}). Once the twin reaches certain critical 
thickness, the crack grows rapidly leading to the brittle failure (Figure \ref{Perfect}d-f). In agreement with the present observation, 
previous reports have shown that the (111)$<$110$>$ crack system leads to brittle failure in BCC Fe \cite{gordon2007crack,al2016atomistic}. 
The brittle failure in this crack orientation is due to the lower energy release rate associated with cleavage ($G_{Cleav})$ than that 
of dislocation nucleation from a crack tip ($G_{Disl})$ i.e., $G_{Cleav} < G_{Disl}$ \cite{gordon2007crack}.

\begin{figure}[h]
\centering
\includegraphics[width=9cm]{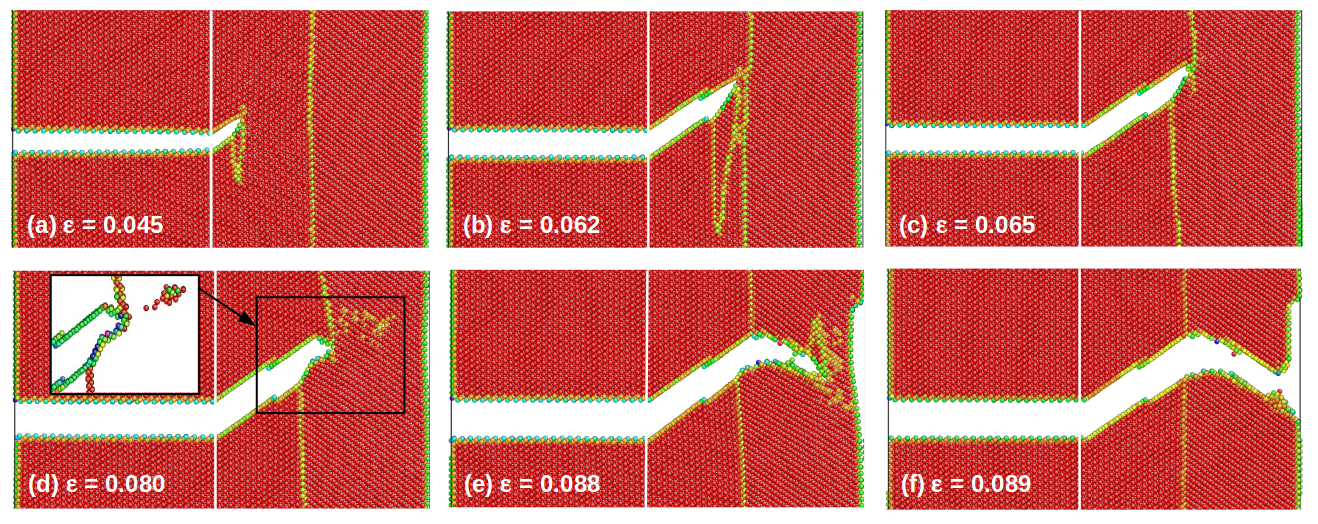}
\caption {\small The atomic configurations demonstrating the growth of an atomistically sharp (111)$<$110$>$ crack in single crystal 
BCC Fe containing a single twin boundary. \hl{The coloring of atoms and the color legend is same as in Fig. 1. The viewing direction 
is $<$110$>$ and only the portion of (110) plane is shown.}}
\label{1-TB}
\end{figure}

Figure \ref{1-TB}(a-f) show the atomic snapshots depicting the growth of (111)$<$110$>$ crack in BCC Fe containing a single TB. Similar 
to perfect crystal, the crack grows in a brittle manner on \{110\} plane with a micro-twin at the crack tip (Figure \ref{1-TB}a-b). 
However, once the crack tip reaches the TB, the micro-twin gets annihilated due to twin-twin interactions (Figure \ref{1-TB}c). The 
annihilation of micro-twin relaxes the stress from 4.8 to 3 GPa (Figure \ref{Stress-strain}). Following the annihilation of micro-twin, 
the 1/2$<$111$>$ full dislocations nucleate continuously from the intersection of crack tip and TB (Figure \ref{1-TB}d) leading to 
significant plastic deformation. The continuous emission of dislocations blunts the crack, temporarily arresting the crack growth. 
Once the strain reaches certain critical value, the crack penetrates the TB and grows rapidly on a symmetric \{110\} plane leading 
to final failure (Figure \ref{1-TB}e-f). Thus, it can be seen that the presence of TB introduces some plastic deformation which blunts 
the crack and in turn delays the brittle failure.  

\begin{figure}[h]
\centering
\includegraphics[width=7cm]{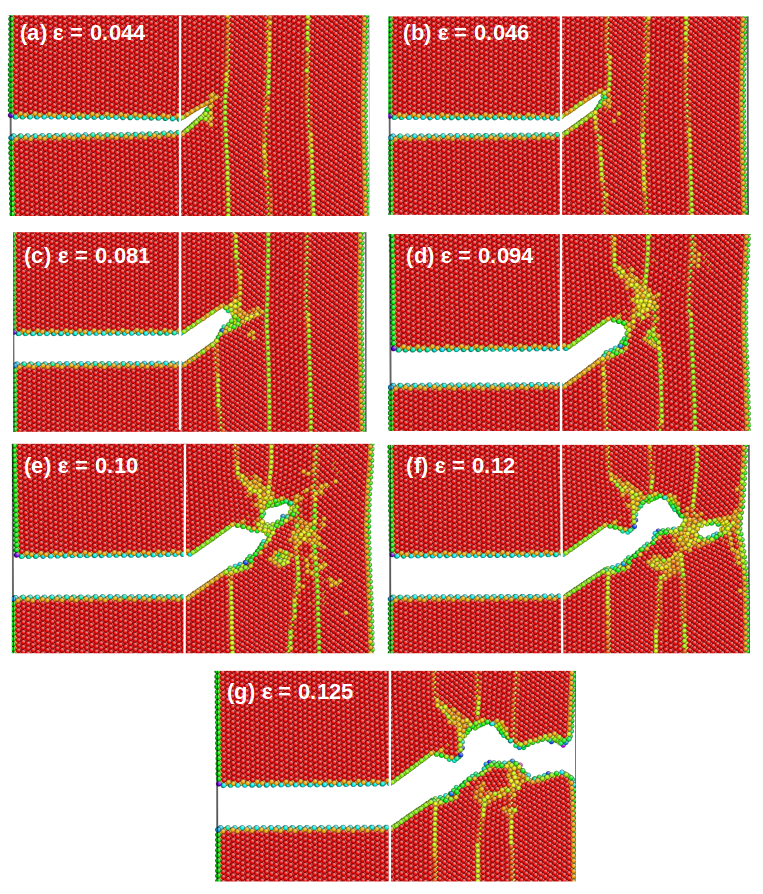}
\caption {\small The atomic configurations depicting the (111)$<$110$>$ crack growth behaviour in single crystal BCC Fe containing 
multiple twin boundaries. \hl{The coloring of atoms and the color legend is same as in Fig. 1. The viewing direction is $<$110$>$ and 
only the portion of (110) plane is shown.}}
\label{3-TB}
\end{figure}

Figure \ref{3-TB} shows the atomic snapshots depicting the growth of (111)$<$110$>$ crack in BCC Fe containing multiple TBs. It can 
be seen that the crack grows in a brittle manner on \{110\} plane without any micro-twin at the crack tip (Figure \ref{3-TB}a-b), 
suggesting that the presence of TBs close to the crack tip may suppress the formation of deformation twin at the tip of the crack. 
Once the crack tip reaches the TB, the crack gets blunted due to the emission of dislocations (Figure \ref{3-TB}c). With increasing 
strain, more and more dislocations are emitted from the intersection of crack tip and TB. As a result, the first TB becomes curved in 
nature and interacts with the middle TB ahead of the crack tip (Figure \ref{3-TB}d). With increasing deformation, a void nucleates 
in the next grain from the twin-twin intersection and merges with the initial crack leading to an increased length of crack (Figure 
\ref{3-TB}e-f). The same process of void nucleation, growth and merging with preceding crack leads to the final failure as shown in 
Figure \ref{3-TB}g. This shows that more the number of TBs more is the plastic deformation, which significantly delays the 
brittle failure. 

\begin{figure}[h]
\centering
\includegraphics[width=6cm]{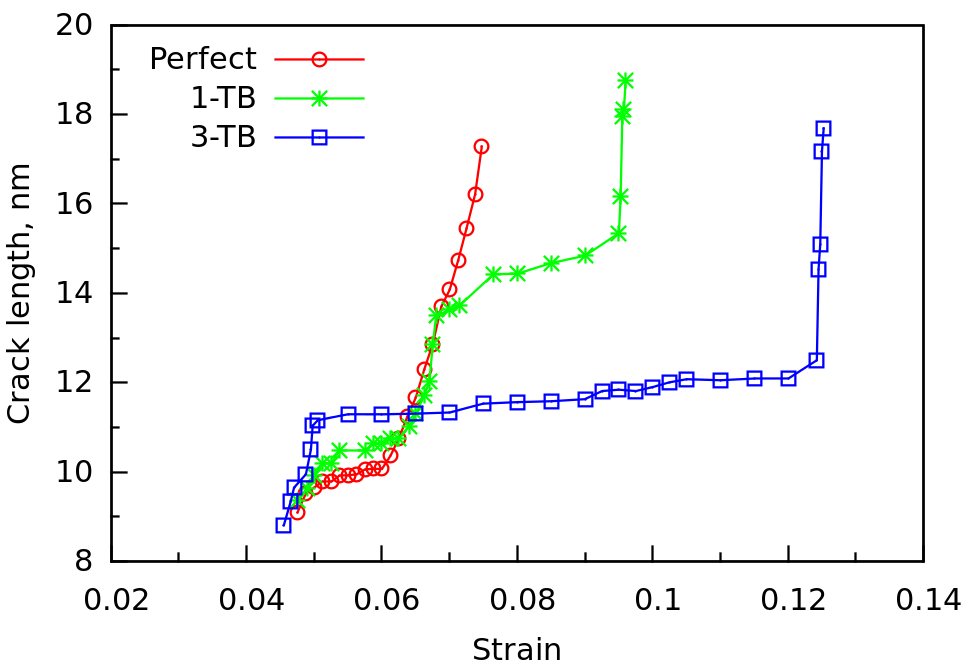}
\caption {\small The crack length as a function of strain during the crack growth behaviour in pre-cracked perfect BCC Fe along 
with that of sample containing single and multiple twin boundaries.}
\label{Crack-length}
\end{figure}

For better understanding the crack growth behaviour, the crack length has been measured as a function of strain. Figure 
\ref{Crack-length} shows the length of the crack as a function of strain in pre-cracked perfect BCC Fe along with twinned 
BCC Fe. \hl{The crack length is measured as the minimum distance between the initial and final positions of a crack tip}.
\hl{In the beginning stage, it can be seen that the crack growth rate increases with increasing the number of TBs until a 
strain value of 0.06. In other words, the initially crack growth rate is high to low for multiple to no TB case. This 
anomalous behaviour may be due to an attractive force of TB on the crack tip. The TBs generally exerts an attractive or 
repulsive force on dislocations and other defects \mbox{\cite{chen2007repulsive,deng2010repulsive,guo2011repulsive,sainath2021role}}.  
As can be seen in Figure \mbox{\ref{Initial}a-c}, in three TBs case, the TB is closure to the crack tip as compared to the 
case of sample with one TB. When the TB is closure to the crack tip, the higher attractive force may be responsible for the  
higher crack growth rate. As the distance between TB and crack tip increases, the attractive force may decreases accordingly, 
thus reducing the crack growth rate in sample with one TB/no TB}. Above the strain value of 0.06, the crack growth rate 
continuously increases in perfect sample until failure, while in sample containing a TB, it slows down \hl{after the strain 
value of 0.07 or the crack length of 13.5 nm}. This indicates that the presence of TB delays the crack growth rate in BCC Fe. 
Once the crack penetrates the TB \hl{at a strain value of 0.095}, once again the crack grows rapidly leading to final failure. 
In contrast to the perfect and single TB sample, the crack grows rapidly in the beginning following which the crack length is 
almost constant until just before failure in sample containing three TBs. This indicates that increasing the number of TBs 
increases the crack growth resistance or decreases the crack growth rate in the sample. 

\hl{The comparison of Figure \mbox{\ref{Crack-length}} with Figure \mbox{\ref{Stress-strain}} suggests} an interesting relation 
between the crack growth rate and flow stress. \hl{It can be seen that}, when the crack grows rapidly (high crack growth rate), 
a sudden drop in flow stress has been noticed. Similarly, when the crack grows gradually (crack growth rate increases linearly), 
the flow stress remains almost constant. On the other hand, when the crack growth rate is very small or negligible, the flow 
stress increases gradually. Similar relation between the flow stress and crack growth rate has been reported in Al \cite{fang2016molecular}.

\section{Conclusions}

The influence of twin boundaries on the growth of an atomically sharp crack has been studied in BCC Fe using molecular 
dynamics simulations. The MD simulation results indicate that in perfect single crystal BCC Fe, the (111)$<$110$>$ crack 
system exhibits a brittle a behaviour, while in system containing TBs, the same crack system results in significant plastic 
deformation, which delays the brittle failure. Further, it has been observed that increasing the number of TBs increases 
the amount of plastic deformation leading to higher crack resistance or lower crack growth rates and higher strain to 
failure. The results on multiple twinned BCC Fe suggest that the presence of TBs close to the crack tip may suppress the 
formation of deformation twin at the crack tip. Finally, an interesting relationship has been observed between the crack 
growth rate and flow stress. When the crack grows rapidly (high crack growth rate), a sudden drop in flow stress is evident. 
Similarly, when the crack grows gradually (crack growth rate is linear), the flow stress remains almost constant. On the 
other hand, when the crack growth rate is very small or negligible, the flow stress increases gradually.

%\begin{acknowledgements}
%If you'd like to thank anyone, place your comments here and remove the percent signs.
%\end{acknowledgements}

% Authors must disclose all relationships or interests that could have direct or potential influence or impart bias on 
\section*{Conflict of interest}

The authors declare that they have no conflict of interest.
\bibliographystyle{spphys}       % APS-like style for physics
\bibliography{my_bibtex-short.bib}   % name your BibTeX data base


% etc

\end{document}